\documentclass[twocolumn]{autart}    

\usepackage{graphicx}          

\usepackage{amssymb}

\usepackage{natbib}

\usepackage{amsfonts}

\usepackage{lipsum}
\usepackage{mathtools}
\usepackage{cuted}

\usepackage{balance}

\usepackage{amsmath}

\DeclareMathOperator*{\esssup}{ess\,sup}

\begin{document}

\begin{frontmatter}

\title{Fundamental Limits of Controlled Stochastic Dynamical Systems: An Information-Theoretic Approach} 


\author[one]{Song Fang}\ead{song.fang@nyu.edu},    
\author[one]{Quanyan Zhu}\ead{quanyan.zhu@nyu.edu}  

\address[one]{Department of Electrical and Computer Engineering, New York University, USA}  


\begin{abstract}                          
In this paper, we examine the fundamental performance limitations in the control of stochastic dynamical systems; more specifically, we derive generic $\mathcal{L}_p$ bounds that hold for any causal (stabilizing) controllers and any stochastic disturbances, by an information-theoretic analysis. We first consider the scenario where the plant (i.e., the dynamical system to be controlled) is linear time-invariant, and it is seen in general that the lower bounds are characterized by the unstable poles (or nonminimum-phase zeros) of the plant as well as the conditional entropy of the disturbance. We then analyze the setting where the plant is assumed to be (strictly) causal, for which case the lower bounds are determined by the conditional entropy of the disturbance. We also discuss the special cases of $p = 2$ and $p = \infty$, which correspond to minimum-variance control and controlling the maximum deviations, respectively. In addition, we investigate the power-spectral characterization of the lower bounds as well as its relation to the Kolmogorov--Szeg\"o formula.  
\end{abstract}

\end{frontmatter}

\section{Introduction}
\label{sec:intro}

Fundamental performance limitation analysis of feedback control systems such as the Bode integral (see, e.g., \citet{Bod:45, aastrom2000limitations, stein2003respect, looze2010, seron2012fundamental} and the references therein) has been of continuing interest throughout classical control theory and modern control theory. In most cases of such performance limitation results, however, specific restrictions on the classes of the controllers that can be implemented must be imposed; one common restriction is that the controllers are assumed to be linear time-invariant (LTI) in the first place \citep{seron2012fundamental}. These restrictions would normally render the analysis invalid in situations where the controllers are allowed to be more general. For instance, learning-based control (based upon, e.g., reinforcement learning and/or deep learning; see \citet{lewis2012reinforcement, mnih2015human, duan2016benchmarking, kocijan2016modelling, duriez2017machine, recht2019tour, bertsekas2019reinforcement, tiumentsev2019neural, zoppoli2020neural, hardt2021patterns} and the references therein) is becoming more and more prevalent nowadays, whereas from an input-output viewpoint the learning algorithms employed are in general not necessarily
LTI.

Information theory, a mathematical theory developed originally for the analysis of fundamental limits of communication systems \citep{shannon1998mathematical, Cov:06}, was in recent years seen to be applicable to the analysis of performance limitations of feedback control systems as well \citep{zang2003nonlinear, Mar:07, Mar:08, Oka:08, Ish:09, Yu:10, Les:10,  hurtado2010limitations, Li:13a, Li:13b, heertjes2013self, ruan2013information, Zha:14, zhao2015effect,  lupu2015information, Fang17Automatica, Fang17TAC, fang2018power, wan2019sensitivity} (see also \citet{fang2017towards, chen2019fundamental} for surveys on this topic).
More specifically, \citet{zang2003nonlinear} obtained a nonlinear extension of Bode's integral based
on an information-theoretic interpretation. In \citet{Mar:07, Mar:08}, an information-theoretic
approach was developed to derive Bode-like integrals, characterizing the fundamental limitations of disturbance attenuation, for feedback control systems, including those in the presence of side information and noisy channels. Subsequently, Bode-like integrals that characterize the complementary sensitivity property as well as  those for multiple-input multiple-output (MIMO) systems were obtained in \citet{Oka:08, Ish:09} via the information-theoretic approach. In addition, this information-theoretic approach has been further employed to derive Bode-like integrals for nonlinear systems \citep{Yu:10}, molecular fluctuations analysis \citep{Les:10}, tracking systems \citep{hurtado2010limitations}, stochastic switched
systems \citep{Li:13a}, continuous-time systems \citep{Li:13b}, master-slave synchronization of high-precision stage systems \citep{heertjes2013self}, vehicle platoon control systems \citep{ruan2013information}, leader-follower systems \citep{Zha:14}, systems with delayed side information about the disturbance \citep{zhao2015effect}, human-machine interaction systems \citep{lupu2015information}, and the characterization of the complementary sensitivity property in continuous-time systems \citep{wan2019sensitivity}; in our previous works, we also leveraged on the information-theoretic approach to develop Bode-like integrals and power gain bounds for networked control systems \citep{Fang17Automatica, Fang17TAC, fang2018power}.
One essential difference between this line of research and the conventional feedback performance limitation analysis is that the information-theoretic performance limits and bounds hold for any causal (stabilizing) controllers, including the aforementioned learning-based controllers as well as LTI controllers as special classes. 
(It is worth pointing out that there also exist other approaches to analyze the performance limitations of feedback control systems while allowing the controllers to be generic; see, e.g., \citet{xie2000much, guo2020feedback} and \citet{nakahira2019connecting,nakahira2020integrative} as well as the references therein.)

In this paper, we go beyond the classes of information-theoretic performance limitations analyzed in the aforementioned works, and analyze the fundamental limits in minimizing the $\mathcal{L}_p$ norms of signals in feedback control systems, by examining the underlying entropic relationships of the signals flowing in the feedback loop.
The fundamental $\mathcal{L}_p$ bounds are shown to hold for any controllers, as deterministic or randomized functions/mappings, as long as they are causal and stabilizing. Meanwhile, the disturbance can be with any distributions; for instance, it is not necessarily independent and identically distributed (i.i.d.), or Gaussian, or even stationary.
In particular, we first consider feedback control systems where the plant is assumed to be LTI while the controller can be generically causal as long as it stabilizes the plant. It is seen that the $\mathcal{L}_p$ norm of the error signal is fundamentally lower bounded by the unstable poles of the plant as well as the conditional entropy of the disturbance, whereas the $\mathcal{L}_p$ norm of the plant output is lowered bounded by nonminimum-phase zeros of the plant together with the disturbance conditional entropy.
We also examine the special cases of $p=2$ and $p=\infty$, and the results reduce to generic lower bounds for minimum-variance control as well as for controlling the maximum deviations, respectively. In addition, we provide a power-spectral characterization of the lower bounds when the disturbance is asymptotically stationary, which establishes the relation to the Kolmogorov--Szeg\"o formula \citep{Pap:02, vaidyanathan2007theory, lindquist2015linear}. 
Finally, we study the case where the plant is also generically assumed to be (strictly) causal.

The remainder of the paper is organized as follows. Section~2 introduces the technical preliminaries. Section~3 presents the main results. 
Concluding remarks are given in Section~4.

Note that this paper is based upon \citet{FangACC21Lp}, which, however, only discusses the case of strictly causal plants (and for only the error signal); such results are essentially what have been presented in Section~3.4 herein. For the current version, we investigate additionally the setting of LTI plants (and for both the error signal and the plant output; see the main result Theorem~3 and Section~3.3), the analysis of which is more sophisticated, as evidenced by the proofs, since it involves combining the entropic analysis and the state-space dynamics. We also include further implications and interpretations of the results (Section~3.1 and Section~3.2). Meanwhile, an arXiv version of this paper can be found in \citet{FangLparxiv}. Note in particular that \citet{FangLparxiv} was titled ``Fundamental Limits on the Maximum Deviations in Control Systems" (focusing on the case of $p = \infty$) for an earlier version, but they are essentially the same paper. This is pointed out herein so as to avoid possible
unnecessary confusions to the readers.

\section{Preliminaries}

In this paper, we consider real-valued continuous random variables and vectors, as well as discrete-time stochastic processes they compose. All random variables, random vectors, and stochastic processes are assumed to be zero-mean. We represent random variables and vectors using boldface letters. Given a stochastic process $\left\{ \mathbf{x}_{k}\right\}$, we denote the sequence $\mathbf{x}_0,\ldots,\mathbf{x}_{k}$ by the random vector $\mathbf{x}_{0,\ldots,k}=\left[\mathbf{x}_0^T~\cdots~\mathbf{x}_{k}^T\right]^T$ for simplicity. The logarithm is defined with base $2$. All functions are assumed to be measurable. Note in particular that, for simplicity and
with abuse of notations, we utilize $\mathbf{x} \in \mathbb{R}$ and $\mathbf{x} \in \mathbb{R}^n$ to
indicate that $\mathbf{x}$ is a real-valued random variable and that $\mathbf{x}$
is a real-valued $n$-dimensional random vector, respectively.

A stochastic process $\left\{ \mathbf{x}_{k}\right\}$ is said to be asymptotically stationary if it is stationary as $k \to \infty$, and herein stationarity means strict stationarity unless otherwise specified \citep{Pap:02}. 
In addition, a process being asymptotically stationary implies that it is asymptotically mean stationary \citep{gray2011entropy}. On the other hand, if a process $\left\{ \mathbf{x}_{k}\right\}, \mathbf{x}_{k} \in \mathbb{R}$, is asymptotically stationary, then its asymptotic autocorrelation 
\begin{flalign} & R_{\mathbf{x}}\left( i,k\right) = 
\lim_{i\to \infty} \mathbb{E}\left[  \mathbf{x}_i \mathbf{x}_{i+k} \right] \nonumber
& \end{flalign}
depends only on $k$, and can thus be denoted as  $R_{\mathbf{x}}\left( k\right)$ for simplicity. Accordingly, the asymptotic power spectrum of such an asymptotically stationary process $\left\{ \mathbf{x}_{k} \right\}$ is defined as
\begin{flalign} &
S_{\mathbf{x}}\left( \omega\right)
=\sum_{k=-\infty}^{\infty} R_{\mathbf{x}}\left( k\right) \mathrm{e}^{-\mathrm{j}\omega k}. \nonumber
& \end{flalign}

Entropy, conditional entropy, and mutual information are the most basic notions in information theory \citep{Cov:06}, which we introduce below.

\begin{defn} The differential entropy of a random vector $\mathbf{x}$ with density $p_{\mathbf{x}} \left(x\right)$ is defined as
	\begin{flalign} &
	h\left( \mathbf{x} \right)
	=-\int p_{\mathbf{x}} \left(x\right) \log p_{\mathbf{x}} \left(x\right) \mathrm{d} x. \nonumber
	& \end{flalign}
	The conditional differential entropy of random vector $\mathbf{x}$ given random vector $\mathbf{y}$ with joint density $p_{\mathbf{x}, \mathbf{y}} \left(x,y\right)$ and conditional density $p_{\mathbf{x} | \mathbf{y}} \left(x,y\right)$ is defined as
	\begin{flalign} &
	h\left(\mathbf{x}\middle|\mathbf{y}\right)
	=-\int p_{\mathbf{x}, \mathbf{y}} \left(x,y\right)\log p_{\mathbf{x} | \mathbf{y}} \left(x,y\right) \mathrm{d}x\mathrm{d}y. \nonumber
	& \end{flalign}
	The mutual information between random vectors $\mathbf{x}, \mathbf{y}$ with densities $p_{\mathbf{x}} \left(x\right)$, $p_{\mathbf{y}} \left( y \right) $ and joint density $p_{\mathbf{x}, \mathbf{y}} \left(x,y\right)$ is defined as
	\begin{flalign} &
	I\left(\mathbf{x};\mathbf{y}\right)
	=\int p_{\mathbf{x}, \mathbf{y}} \left(x,y\right) \log \frac{p_{\mathbf{x}, \mathbf{y}} \left(x,y\right)}{p_{\mathbf{x}} \left(x\right) p_{\mathbf{y}} \left( y \right) }\mathrm{d}x\mathrm{d}y. \nonumber
	& \end{flalign}
	The entropy rate of a stochastic process $\left\{ \mathbf{x}_{k}\right\}$ is defined as
	\begin{flalign} &
	h_\infty \left(\mathbf{x}\right)=\limsup_{k\to \infty} \frac{h\left(\mathbf{x}_{0,\ldots,k}\right)}{k+1}. \nonumber
	& \end{flalign}
\end{defn}


Properties of these notions can be found in, e.g., \citet{Cov:06}. In particular, the next lemma \citep{dolinar1991maximum} presents the maximum-entropy probability distributions under $\mathcal{L}_{p}$-norm constraints for random variables.

\begin{lem} \label{maximum}
	Consider a random variable $\mathbf{x} \in \mathbb{R}$ with $\mathcal{L}_{p}$ norm $\left[ \mathbb{E} \left( \left| \mathbf{x} \right|^{p} \right) \right]^{\frac{1}{p}} = \mu,~p \geq 1$.
	Then,  
	\begin{flalign} & 
	h \left( \mathbf{x} \right) 
	\leq \log \left[ 2 \Gamma \left( \frac{p+1}{p} \right) \left( p \mathrm{e} \right)^{\frac{1}{p}} \mu \right], \nonumber 
	& \end{flalign}
	where equality holds if and only if $\mathbf{x}$ is with probability density function
	\begin{flalign} &
	f_{\mathbf{x}} \left( x \right)
	= \frac{ \mathrm{e}^{- \left| x \right|^{p} / \left( p \mu^{p} \right)} }{2 \Gamma \left( \frac{p+1}{p} \right) p^{\frac{1}{p}} \mu}. \nonumber
	& \end{flalign}
	Herein, $\Gamma \left( \cdot \right)$ denotes the Gamma function.
\end{lem}

In particular, when $p \to \infty$, 
\begin{flalign} &
\lim_{p \to \infty} \left[ \mathbb{E} \left( \left| \mathbf{x} \right|^{p} \right) \right]^{\frac{1}{p}} = \esssup_{ f_{\mathbf{x}} \left( x \right) > 0} \left| \mathbf{x} \right|, \nonumber
& \end{flalign}
and
\begin{flalign} & 
\lim_{p \to \infty} \log \left[ 2 \Gamma \left( \frac{p+1}{p} \right) \left( p \mathrm{e} \right)^{\frac{1}{p}} \mu \right] = \log \left( 2 \mu \right), \nonumber
& \end{flalign}
while
\begin{flalign} &
\lim_{p \to \infty}\frac{ \mathrm{e}^{- \left| x \right|^{p} / \left( p \mu^{p} \right)} }{2 \Gamma \left( \frac{p+1}{p} \right) p^{\frac{1}{p}} \mu}
= 
\left\{ \begin{array}{cc}
\frac{1}{2 \mu}, & \left| x \right| \leq \mu,\\
0, & \left| x \right| > \mu.
\end{array} \right. \nonumber
& \end{flalign}

%

\section{Generic $\mathcal{L}_p$ Bounds in Feedback Control Systems}

In this section, we examine how the minimization of $\mathcal{L}_p$ norms of signals in feedback control systems is fundamentally limited by the properties of the plant and disturbance, no matter what controllers are to be utilized as long as they are causal and stabilizing.


%


\begin{figure}
	\begin{center}
		\vspace{-3mm}
		\includegraphics [width=0.4\textwidth]{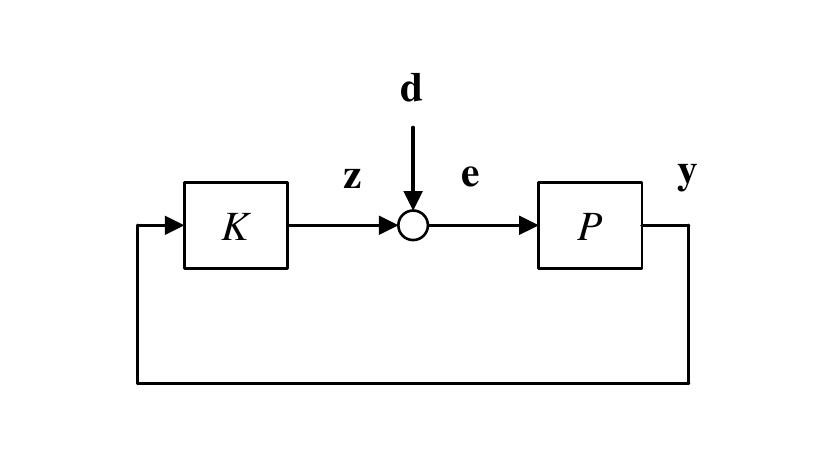}
		\vspace{-6mm}
		\caption{A feedback control system.}
		\label{feedback}
	\end{center}
	\vspace{6mm}
\end{figure}

We first consider the feedback control system depicted in Fig.~\ref{feedback}. Herein, the plant $P$ is assumed to be LTI with state-space model given by
\begin{flalign} & \label{plant}
\left\{ \begin{array}{rcl}
\mathbf{x}_{k+1} & = & A \mathbf{x}_{k} + B \mathbf{e}_{k},\\
\mathbf{y}_{k} & = & C \mathbf{x}_{k},
\end{array} \right. 
& \end{flalign}
where $\mathbf{x}_{k} \in \mathbb{R}^n$ is the plant state, $\mathbf{e}_{k} \in \mathbb{R}$ is the plant input, and $\mathbf{y}_{k} \in \mathbb{R}$ is the plant output.
The system matrices are $ A \in \mathbb{R}^{n \times n}$, $ B \in \mathbb{R}^{n \times 1}$, and $ C \in \mathbb{R}^{1 \times n}$. 

Meanwhile, the controller $K$ is generically assumed to be causal, i.e., for any time instant $k \geq 0$, 
\begin{flalign} & \label{controller}
\mathbf{z}_k = K_{k} \left( \mathbf{y}_{0,\ldots,k} \right),
& \end{flalign}
where the plant output $\mathbf{y}_{k} \in \mathbb{R}$ is now the controller input (through feedback) while $\mathbf{z}_{k} \in \mathbb{R}$ is the controller output. Note in particular that herein $K_{k} \left( \cdot \right)$ may represent any deterministic or randomized functions/mappings. This is a very general assumption in the sense that the controller can be linear or nonlinear, time-invariant or time-varying, and so on, as long as it is causal and stabilizing, whereas we say that the controller $K$ stabilizes the plant $P$ if 
\begin{flalign} &
\lim_{k \to \infty} \mathbb{E} \left[  \mathbf{x}_k^{\mathrm{T}} \mathbf{x}_k \right] < \infty, 
& \end{flalign}
i.e., the closed-loop system is asymptotically mean-square stable. This assumption essentially allows all possible controllers that can be realized physically in practical use.

Furthermore, suppose that an additive disturbance $\left\{ \mathbf{d}_{k} \right\}$ exists between the controller output $\left\{ \mathbf{z}_{k} \right\}$ and plant input $\left\{ \mathbf{e}_{k} \right\}$, that is,
\begin{flalign} &
\mathbf{e}_{k} = \mathbf{z}_{k} + \mathbf{d}_{k}.
& \end{flalign}
Meanwhile, it is assumed that $\left\{ \mathbf{d}_{k} \right\}$, $\mathbf{x}_0$, and $\mathbf{z}_0$ are mutually independent.

For such a feedback control system, the following asymptotic bound on the $\mathcal{L}_p$ norm of the error signal always holds.

\begin{thm} \label{t1}
	Consider the control system depicted in Fig.~\ref{feedback}, where the plant $P$ is given by \eqref{plant} while the controller $K$ is given by \eqref{controller}. If $K$ stabilizes $P$, then the $\mathcal{L}_p$ norm of $ \mathbf{e}_{k} $ is asymptotically lower bounded by 
	\begin{flalign} & \label{bound}
	\limsup_{k \to \infty} \left[ \mathbb{E} \left( \left| \mathbf{e}_{k} \right|^{p} \right) \right]^{\frac{1}{p}} \nonumber \\
	&~~~~ \geq \frac{1}{2 \Gamma \left( \frac{p+1}{p} \right) \left( p \mathrm{e} \right)^{\frac{1}{p}}} \left[\prod_{i=1}^{n} \max \left\{1, \left| \lambda_{i} \left( A \right) \right|  \right\} \right] \nonumber \\
	&~~~~~~~~ \times 2^{\limsup_{k \to \infty} h \left( \mathbf{d}_{k} |  \mathbf{d}_{0,\ldots,k-1} \right) },
	& \end{flalign}
	where $\lambda_i \left( A \right), i = 1, \ldots, m$, denote the eigenvalues of $A$, while $\limsup_{k \to \infty} h \left( \mathbf{d}_{k} |  \mathbf{d}_{0,\ldots,k-1} \right)$ denotes the asymptotic conditional entropy of $\mathbf{d}_k$ given $\mathbf{d}_{0,\ldots,k-1} $.
\end{thm}


\begin{pf}
	We shall first prove the fact that $\mathbf{z}_{k}$ is eventually a function of $\mathbf{d}_{0,\ldots,k-1}$, $\mathbf{z}_{0}$, and $\mathbf{x}_{0}$.
	Note that when $k=0$, \eqref{controller} reduces to  
	\begin{flalign} &
	\mathbf{z}_{0} = K_{0} \left( \mathbf{y}_{0} \right) = K_{0} \left( C \mathbf{x}_{0} \right), \nonumber
	& \end{flalign}
	that is, $\mathbf{z}_{0}$ is a function of $\mathbf{x}_{0}$.
	Next, when $k=1$, 
	\eqref{controller} is given by 
	\begin{flalign} 
	\mathbf{z}_{1} 
	& = K_{1} \left( \mathbf{y}_{0}, \mathbf{y}_{1} \right) = K_{1} \left( C \mathbf{x}_{0}, C \mathbf{x}_{1} \right) \nonumber \\
	&= K_{1} \left( C \mathbf{x}_{0}, C \left( A \mathbf{x}_{0} + B \mathbf{e}_{0} \right) \right). \nonumber
	& \end{flalign}
	In other words, noting also that $\mathbf{e}_{0} = \mathbf{z}_{0} + \mathbf{d}_{0}$, $\mathbf{z}_{1}$ is a function of $\mathbf{d}_{0}$, $\mathbf{z}_{0}$, and $\mathbf{x}_{0}$. In addition, 
	when $k=2$, 
	\eqref{controller} is given by 
	\begin{flalign} 
	&\mathbf{z}_{2} 
	= K_{2} \left( \mathbf{y}_{0}, \mathbf{y}_{1}, \mathbf{y}_{2} \right) = K_{2} \left( C \mathbf{x}_{0}, C \mathbf{x}_{1}, C \mathbf{x}_{2} \right) \nonumber \\
	&= K_{2} \left( C \mathbf{x}_{0}, C \left( A \mathbf{x}_{0} + B \mathbf{e}_{0} \right), C \left( A \mathbf{x}_{1} + B \mathbf{e}_{1} \right) \right)
	\nonumber \\
	&= K_{2} \left( C \mathbf{x}_{0}, C \left( A \mathbf{x}_{0} + B \mathbf{e}_{0} \right), C \left[ A \left( A \mathbf{x}_{0} + B \mathbf{e}_{0} \right) + B \mathbf{e}_{1} \right] \right). \nonumber
	& \end{flalign}
	That is to say, $\mathbf{z}_{2}$ is a function of $\mathbf{d}_{0,1}$, $\mathbf{z}_{0}$, and $\mathbf{x}_{0}$, noting as well that $\mathbf{e}_{0} = \mathbf{z}_{0} + \mathbf{d}_{0}$ and $\mathbf{e}_{1} = \mathbf{z}_{1} + \mathbf{d}_{1}$ whereas we have previously proved that $\mathbf{z}_{1}$ is a function of $\mathbf{d}_{0}$, $\mathbf{z}_{0}$, and $\mathbf{x}_{0}$.
	We may then repeat this process and verify that for any $k \geq 0$, $\mathbf{z}_{k}$ is eventually a function of $\mathbf{d}_{0,\ldots,k-1}$, $\mathbf{z}_{0}$, and $\mathbf{x}_{0}$.
	
	Having shown this fact, we will then proceed to prove the main result of this theorem. To begin with, it follows from Lemma~\ref{maximum} that
	\begin{flalign} &
	\left[ \mathbb{E} \left( \left| \mathbf{e}_{k} \right|^{p} \right) \right]^{\frac{1}{p}}
	\geq \frac{2^{h \left(  \mathbf{e}_{k} \right)}}{2 \Gamma \left( \frac{p+1}{p} \right) \left( p \mathrm{e} \right)^{\frac{1}{p}}}, \nonumber
	& \end{flalign}
	where equality holds
	if and only if $\mathbf{e}_{k}$ is with probability density function
	\begin{flalign} &
	f_{\mathbf{e}_{k}} \left( x \right)
	= \frac{ \mathrm{e}^{- \left| x \right|^{p} / \left( p \mu^{p} \right)} }{2 \Gamma \left( \frac{p+1}{p} \right) p^{\frac{1}{p}} \mu}, \nonumber
	& \end{flalign}
	whereas
	\begin{flalign} &
	\mu 
	= \frac{2^{h \left( \mathbf{e}_{k} \right)}}{2 \Gamma \left( \frac{p+1}{p} \right) \left( p \mathrm{e} \right)^{\frac{1}{p}}}. \nonumber
	& \end{flalign}
	Meanwhile,
	\begin{flalign} 
	h \left( \mathbf{e}_{k} \right)
	& = h \left( \mathbf{e}_{k} |  \mathbf{d}_{0,\ldots,k-1}, \mathbf{z}_{0}, \mathbf{x}_{0} \right) + I \left( \mathbf{e}_{k}; \mathbf{d}_{0,\ldots,k-1}, \mathbf{z}_{0}, \mathbf{x}_{0} \right) \nonumber \\
	& = h \left( \mathbf{z}_{k} + \mathbf{d}_{k} |  \mathbf{d}_{0,\ldots,k-1}, \mathbf{z}_{0}, \mathbf{x}_{0} \right) \nonumber \\
	&~~~~ + I \left( \mathbf{e}_{k}; \mathbf{d}_{0,\ldots,k-1}, \mathbf{z}_{0}, \mathbf{x}_{0} \right). \nonumber
	& \end{flalign}
	Then, according to the fact that $\mathbf{z}_{k}$ is a function of $\mathbf{d}_{0,\ldots,k-1}$, $\mathbf{z}_{0}$, and $\mathbf{x}_{0}$, we have
	\begin{flalign} & 
	h \left( \mathbf{z}_{k} + \mathbf{d}_{k} |  \mathbf{d}_{0,\ldots,k-1}, \mathbf{z}_{0}, \mathbf{x}_{0} \right) = h \left( \mathbf{d}_{k} |  \mathbf{d}_{0,\ldots,k-1}, \mathbf{z}_{0}, \mathbf{x}_{0} \right). \nonumber
	& \end{flalign}
	On the other hand, since $ \left\{ \mathbf{d}_{k} \right\}$ is independent of $\mathbf{z}_{0}$ and $\mathbf{x}_{0}$ (and thus $\mathbf{d}_{k}$ is independent of $\mathbf{z}_{0}$ and $\mathbf{x}_{0}$ given $\mathbf{d}_{0,\ldots,k-1}$), we have
	\begin{flalign} 
	&h \left( \mathbf{d}_{k} |  \mathbf{d}_{0,\ldots,k-1}, \mathbf{z}_{0}, \mathbf{x}_{0} \right) \nonumber \\
	&~~~~ = h \left( \mathbf{d}_{k} |  \mathbf{d}_{0,\ldots,k-1} \right) - I \left( \mathbf{d}_{k}; \mathbf{z}_{0}, \mathbf{x}_{0} |  \mathbf{d}_{0,\ldots,k-1} \right) \nonumber \\
	&~~~~  = h \left( \mathbf{d}_{k} |  \mathbf{d}_{0,\ldots,k-1} \right). \nonumber
	& \end{flalign}
	As a result,
	\begin{flalign} 
	h \left( \mathbf{e}_{k} \right)
	&= h \left( \mathbf{d}_{k} |  \mathbf{d}_{0,\ldots,k-1} \right) + I \left( \mathbf{e}_{k}; \mathbf{d}_{0,\ldots,k-1}, \mathbf{z}_{0}, \mathbf{x}_{0} \right). \nonumber
	& \end{flalign}
	Thus,
	\begin{flalign} 
	&\limsup_{k \to \infty} h \left( \mathbf{e}_{k} \right) \nonumber \\
	&~~~~ = \limsup_{k \to \infty} \left[ h \left( \mathbf{d}_{k} |  \mathbf{d}_{0,\ldots,k-1} \right) +  I \left( \mathbf{e}_{k}; \mathbf{d}_{0,\ldots,k-1}, \mathbf{z}_{0}, \mathbf{x}_{0} \right) \right] \nonumber \\
	&~~~~ \geq \limsup_{k \to \infty} h \left( \mathbf{d}_{k} |  \mathbf{d}_{0,\ldots,k-1} \right) \nonumber \\
	&~~~~~~~~ + \liminf_{k \to \infty}  I \left( \mathbf{e}_{k}; \mathbf{d}_{0,\ldots,k-1}, \mathbf{z}_{0}, \mathbf{x}_{0} \right). \nonumber
	& \end{flalign}
	On the other hand,
	\begin{flalign} 
	&\liminf_{k \to \infty} I \left(\mathbf{e}_k; \mathbf{d}_{0,\ldots,k-1}, \mathbf{z}_0, \mathbf{x}_0 \right) \nonumber \\
	&~~~~ = \liminf_{k \to \infty} \frac{I \left( \mathbf{e}_0; \mathbf{z}_0, \mathbf{x}_0 \right) + \cdots + I \left(\mathbf{e}_k; \mathbf{d}_{0,\ldots,k-1}, \mathbf{z}_0, \mathbf{x}_0 \right) }{k+1}. \nonumber
	& \end{flalign}
	Meanwhile, note that $\mathbf{e}_{k-1}$ is eventually a function of $\mathbf{d}_{0,\ldots,k-1}$, $\mathbf{z}_0$, and $ \mathbf{x}_0$, which follows from the fact that $\mathbf{z}_{k-1}$ is a function of $\mathbf{d}_{0,\ldots,k-1}$, $\mathbf{z}_0$, and $ \mathbf{x}_0$ whereas 
	\begin{flalign}
	&\mathbf{e}_{k-1} = \mathbf{z}_{k-1} + \mathbf{d}_{k-1}. \nonumber
	& \end{flalign}
	As such, $\mathbf{e}_{0,\ldots,k-1}$ is also eventually a function of $\mathbf{d}_{0,\ldots,k-1}$, $\mathbf{z}_0$, and $ \mathbf{x}_0$. Hence,
	\begin{flalign}
	& I \left(\mathbf{e}_k; \mathbf{d}_{0,\ldots,k-1}, \mathbf{z}_0, \mathbf{x}_0 \right) = I \left(\mathbf{e}_k; \mathbf{e}_{0,\ldots,k-1}, \mathbf{d}_{0,\ldots,k-1}, \mathbf{z}_0, \mathbf{x}_0 \right). \nonumber
	& \end{flalign}
	In addition, it holds that
	\begin{flalign}
	& I \left( \mathbf{e}_k; \mathbf{e}_{0,\ldots,k-1}, \mathbf{d}_{0,\ldots,k-1}, \mathbf{z}_0, \mathbf{x}_0 \right) \nonumber \\
	& = I \left( \mathbf{e}_k; \mathbf{e}_{0,\ldots,k-1}, \mathbf{x}_0 \right) +  I \left( \mathbf{e}_k;  \mathbf{d}_{0,\ldots,k-1}, \mathbf{z}_0 | \mathbf{e}_{0,\ldots,k-1}, \mathbf{x}_0 \right) \nonumber \\
	& \geq  I \left( \mathbf{e}_k; \mathbf{e}_{0,\ldots,k-1}, \mathbf{x}_0 \right), \nonumber
	& \end{flalign}
	while
	\begin{flalign}
	&I \left(\mathbf{e}_k; \mathbf{e}_{0,\ldots,k-1}, \mathbf{x}_0 \right) \nonumber \\
	&~~~~ = I \left(\mathbf{e}_k; \mathbf{x}_0|\mathbf{e}_{0,\ldots,k-1} \right) + I \left(\mathbf{e}_k; \mathbf{e}_{0,\ldots,k-1} \right) \nonumber \\ 
	&~~~~ \geq I \left(\mathbf{e}_k; \mathbf{x}_0|\mathbf{e}_{0,\ldots,k-1} \right).\nonumber
	& \end{flalign}
	As a result,
	\begin{flalign}
	&I \left(\mathbf{e}_k; \mathbf{d}_{0,\ldots,k-1}, \mathbf{z}_0, \mathbf{x}_0 \right) \geq I \left(\mathbf{e}_k; \mathbf{x}_0|\mathbf{e}_{0,\ldots,k-1} \right),\nonumber
	& \end{flalign}
	or more specifically,
	\begin{flalign} &
	\left\{ \begin{array}{rcl}
	I \left(\mathbf{e}_0; \mathbf{z}_0, \mathbf{x}_0\right) & \geq & I \left( \mathbf{e}_{0}; \mathbf{x}_{0} \right),\\
	I \left(\mathbf{e}_1; \mathbf{d}_{0}, \mathbf{z}_0, \mathbf{x}_0\right) & \geq & I \left( \mathbf{e}_{1}; \mathbf{x}_{0} | \mathbf{e}_{0} \right),
	\\
	& \vdots & 
	\\
	I \left(\mathbf{e}_k; \mathbf{d}_{0,\ldots,k-1}, \mathbf{z}_0, \mathbf{x}_0\right) & \geq & I \left( \mathbf{e}_{k}; \mathbf{x}_{0} | \mathbf{e}_{0,\ldots,k-1} \right).
	\end{array} \right. \nonumber
	& \end{flalign}
	Accordingly,
	\begin{flalign} 
	&I \left(\mathbf{e}_0; \mathbf{z}_0, \mathbf{x}_0\right) + I \left(\mathbf{e}_1; \mathbf{d}_0, \mathbf{z}_0, \mathbf{x}_0\right) + \cdots \nonumber \\
	&~~~~+ I \left(\mathbf{e}_k; \mathbf{d}_{0,\ldots,k-1}, \mathbf{z}_0, \mathbf{x}_0\right) \nonumber \\
	& \geq I \left( \mathbf{e}_{0}; \mathbf{x}_{0} \right) + I \left( \mathbf{e}_{1}; \mathbf{x}_{0} | \mathbf{e}_{0} \right) + \cdots + I \left( \mathbf{e}_{k}; \mathbf{x}_{0} | \mathbf{e}_{0,\ldots,k-1} \right) \nonumber \\
	& = I \left( \mathbf{e}_{0, \ldots, k}; \mathbf{x}_{0} \right). \nonumber
	& \end{flalign}
	Then, noting as well that \citep{Mar:07}
	\begin{flalign} & 
	\liminf_{k \to \infty} \frac{I \left( \mathbf{e}_{0, \ldots, k}; \mathbf{x}_{0} \right)}{k+1}
	\geq \sum_{i=1}^{n} \max \left\{0, \log \left| \lambda_{i} \left( A \right) \right|  \right\}, \nonumber
	& \end{flalign}
	we have
	\begin{flalign} 
	&\liminf_{k \to \infty}  I \left( \mathbf{e}_{k}; \mathbf{d}_{0,\ldots,k-1}, \mathbf{z}_{0}, \mathbf{x}_{0} \right)  \geq \liminf_{k \to \infty} \frac{I \left( \mathbf{e}_{0, \ldots, k}; \mathbf{x}_{0} \right)}{k+1} \nonumber \\
	&~~~~ \geq \sum_{i=1}^{n} \max \left\{0, \log \left| \lambda_{i} \left( A \right) \right|  \right\}, \nonumber
	& \end{flalign}
	and thus
	\begin{flalign} 
	\limsup_{k \to \infty} h \left( \mathbf{e}_{k} \right)
	&\geq \limsup_{k \to \infty} h \left( \mathbf{d}_{k} |  \mathbf{d}_{0,\ldots,k-1} \right) \nonumber \\
	&~~~~ + \sum_{i=1}^{n} \max \left\{0, \log \left| \lambda_{i} \left( A \right) \right|  \right\}. \nonumber
	& \end{flalign}
	Accordingly,
	\begin{flalign} 
	&\limsup_{k \to \infty} \left[ \mathbb{E} \left( \left| \mathbf{e}_{k} \right|^{p} \right) \right]^{\frac{1}{p}} 
	\geq \limsup_{k \to \infty} \frac{1}{2 \Gamma \left( \frac{p+1}{p} \right) \left( p \mathrm{e} \right)^{\frac{1}{p}}} 2^{h \left(  \mathbf{e}_{k} \right) } \nonumber \\
	&~~~~ = \frac{1}{2 \Gamma \left( \frac{p+1}{p} \right) \left( p \mathrm{e} \right)^{\frac{1}{p}}} 2^{\limsup_{k \to \infty} h \left(  \mathbf{e}_{k} \right) } \nonumber \\
	&~~~~ \geq \frac{1}{2 \Gamma \left( \frac{p+1}{p} \right) \left( p \mathrm{e} \right)^{\frac{1}{p}}} \nonumber \\
	&~~~~~~~~ \times 2^{\limsup_{k \to \infty} h \left( \mathbf{d}_{k} |  \mathbf{d}_{0,\ldots,k-1} \right) + \sum_{i=1}^{n} \max \left\{0, \log \left| \lambda_{i} \left( A \right) \right|  \right\}}
	\nonumber \\
	&~~~~ = \frac{1}{2 \Gamma \left( \frac{p+1}{p} \right) \left( p \mathrm{e} \right)^{\frac{1}{p}}} \left[\prod_{i=1}^{n} \max \left\{1, \left| \lambda_{i} \left( A \right) \right|  \right\} \right] \nonumber \\
	&~~~~~~~~ \times 2^{\limsup_{k \to \infty} h \left( \mathbf{d}_{k} |  \mathbf{d}_{0,\ldots,k-1} \right) }. \nonumber
	& \end{flalign}
	This completes the proof.
\qed \end{pf}

It is worth mentioning that in Theorem~\ref{t1}, no specific restrictions have been imposed on the distribution of the disturbance $\left\{ \mathbf{d}_{k} \right\}$; for instance, it is not necessarily i.i.d. or Gaussian. Meanwhile, the disturbance $\left\{ \mathbf{d}_{k} \right\}$ and the error signal $\left\{ \mathbf{e}_{k} \right\}$ are not required to be stationary or asymptotically stationary either.

On the right-hand side of \eqref{bound}, the term 
\begin{flalign} &
\prod_{i=1}^{n} \max \left\{1, \left| \lambda_{i} \left( A \right) \right|  \right\} 
& \end{flalign}
is essentially the product of (the magnitudes of) all the unstable poles of the plant, which quantifies its degree of instability. Meanwhile, 
\begin{flalign} &
\limsup_{k \to \infty} h \left( \mathbf{d}_{k} |  \mathbf{d}_{0,\ldots,k-1} \right) 
& \end{flalign}
denotes the asymptotic conditional entropy of the current disturbance $\mathbf{d}_{k}$ given the previous disturbances $\mathbf{d}_{0,\ldots,k-1}$, which may be viewed as a measure of randomness contained in $\mathbf{d}_{k}$ given $\mathbf{d}_{0,\ldots,k-1}$ as $k \to \infty$. 
In particular, if $\left\{ \mathbf{d}_{k} \right\}$ is an asymptotically Markov process, then \citep{Cov:06}
\begin{flalign} & 
\limsup_{k \to \infty} h \left( \mathbf{d}_k | \mathbf{d}_{0,\ldots,k-1} \right) = \limsup_{k \to \infty} h \left( \mathbf{d}_k | \mathbf{d}_{k-1} \right).
& \end{flalign}
Additionally, in the extreme case when $\left\{ \mathbf{d}_{k} \right\}$ is asymptotically white, we have 
\begin{flalign} &
\limsup_{k \to \infty} h \left( \mathbf{d}_k | \mathbf{d}_{0,\ldots,k-1} \right) = \limsup_{k \to \infty} h \left( \mathbf{d}_k \right).
& \end{flalign}
Meanwhile, if $\left\{ \mathbf{d}_{k} \right\}$ is assumed to be asymptotically stationary, then it holds that \citep{Cov:06}
\begin{flalign} \label{entropyrate}
\limsup_{k \to \infty} h \left( \mathbf{d}_{k} |  \mathbf{d}_{0,\ldots,k-1} \right)
&= \lim_{k \to \infty} h \left( \mathbf{d}_{k} |  \mathbf{d}_{0,\ldots,k-1} \right) \nonumber \\
&= h_{\infty} \left( \mathbf{d} \right).
& \end{flalign}

In fact, it is known from the proof of Theorem~\ref{t1} that one necessary condition for achieving the lower bound in \eqref{bound} is that $\mathbf{e}_k$ is with probability density function
\begin{flalign} \label{distribution} &
f_{\mathbf{e}_{k}} \left( x \right)
= \frac{ \mathrm{e}^{- \left| x \right|^{p} / \left( p \mu^{p} \right)} }{2 \Gamma \left( \frac{p+1}{p} \right) p^{\frac{1}{p}} \mu}, 
& \end{flalign}
where
\begin{flalign} &
\mu 
= \frac{2^{h \left( \mathbf{e}_{k} \right)}}{2 \Gamma \left( \frac{p+1}{p} \right) \left( p \mathrm{e} \right)^{\frac{1}{p}}}.
& \end{flalign}
This means that if the disturbance $\left\{ \mathbf{d}_k \right\}$ is, e.g., Gaussian, then the optimal controller (in the sense of minimizing the $\mathcal{L}_p$ norm of $\mathbf{e}_k$ asymptotically) must be nonlinear when $p \neq 2$. Otherwise, with a linear controller, $\mathbf{e}_k$ will also be Gaussian, considering that the plant is linear and thus the feedback system is linear as well; note that \eqref{distribution} represents the Gaussian distribution if and only if $p = 2$, and thus a Gaussian $\mathbf{e}_k$ indicates that the controller is not optimal when $p \neq 2$.

\subsection{Special Cases} \label{special}

We now consider the special cases of Theorem~\ref{t1} for when $p=2$ and $p=\infty$, respectively.

%

\subsubsection{When $p=2$} \label{variancesection}
The next corollary follows when $p=2$.

\begin{cor}
	Consider the control system depicted in Fig.~\ref{feedback}, where the plant $P$ is given by \eqref{plant} while the controller $K$ is given by \eqref{controller}. If $K$ stabilizes $P$, then
	\begin{flalign} \label{bound8}
	&\limsup_{k \to \infty} \left[ \mathbb{E} \left( \left| \mathbf{e}_{k} \right|^{2} \right) \right]^{\frac{1}{2}} \nonumber \\
	&\geq \frac{1}{ \sqrt{2 \pi \mathrm{e}}} \left[\prod_{i=1}^{n} \max \left\{1, \left| \lambda_{i} \left( A \right) \right|  \right\} \right] 2^{\limsup_{k \to \infty} h \left( \mathbf{d}_{k} |  \mathbf{d}_{0,\ldots,k-1} \right) },
	& \end{flalign}
\end{cor}

It is clear that \eqref{bound8} can simply be rewritten as
\begin{flalign}
&\limsup_{k \to \infty} \mathbb{E} \left( \mathbf{e}_{k}^{2} \right) \nonumber \\
&\geq \frac{1}{ 2 \pi \mathrm{e}} \left[\prod_{i=1}^{n} \max \left\{1, \left| \lambda_{i} \left( A \right) \right|  \right\} \right]^2 2^{2 \limsup_{k \to \infty} h \left( \mathbf{d}_{k} |  \mathbf{d}_{0,\ldots,k-1} \right) },
& \end{flalign}
which provides a fundamental lower bound for minimum-variance control \citep{aastrom2012introduction}.

\subsubsection{When $p=\infty$} The following result holds when $p=\infty$.

\begin{cor} 
	Consider the control system depicted in Fig.~\ref{feedback}, where the plant $P$ is given by \eqref{plant} while the controller $K$ is given by \eqref{controller}. If $K$ stabilizes $P$, then
	\begin{flalign} \label{bound9}
	&\limsup_{k \to \infty} \esssup_{ f_{\mathbf{e}_{k}} \left( x \right) > 0} \left| \mathbf{e}_{k} \right| \nonumber \\
	&~~~~ \geq \frac{1}{ 2 } \left[\prod_{i=1}^{n} \max \left\{1, \left| \lambda_{i} \left( A \right) \right|  \right\} \right] 2^{\limsup_{k \to \infty} h \left( \mathbf{d}_{k} |  \mathbf{d}_{0,\ldots,k-1} \right) },
	& \end{flalign}
\end{cor}


Note that herein $\esssup_{ f_{\mathbf{e}_{k}} \left( x \right) > 0} \left| \mathbf{e}_{k} \right|$ represents the maximum (worst-case) absolute deviation of $ \mathbf{e}_{k} $ from its mean $\mathbb{E} \left[ \mathbf{e}_{k} \right]$, which is assumed to be zero by default and hence $ \left| \mathbf{e}_{k} \right| = \left| \mathbf{e}_{k} - \mathbb{E} \left[ \mathbf{e}_{k} \right] \right|$. (In fact, the essential supremum gives the smallest positive number that upper bounds the deviation almost surely.)


It is worth pointing out that in the case where the variance of the error is minimized (see Section~\ref{variancesection}), it is possible that the probability of having an arbitrary large deviation (from the mean) in the error signal is non-zero. More specifically, it is known from the proof of Theorem~\ref{t1} that one necessary condition for achieving the lower bound in \eqref{bound8} is that $\mathbf{e}_k$ is with probability density function (corresponding to $p = 2$)
\begin{flalign} &
f_{\mathbf{e}_{k}} \left( x \right)
= \frac{ \mathrm{e}^{- x^2 / \left( 2 \mu^2 \right)} }{ \sqrt{2 \pi} \mu}, 
& \end{flalign}
which represents a Gaussian distribution. This implicates that the probability of having an arbitrary large deviation in the error signal is non-zero, as a consequence of the property of Gaussian distributions,
which could cause severe consequences in safety-critical systems interacting with real world, especially in scenarios where worst-case performance
guarantees must be strictly imposed. Instead, we may directly consider the worst-case scenario by minimizing the maximum deviation rather than the variance of the error in the first place. Accordingly, \eqref{bound9} provides a generic lower bound for minimizing the maximum deviation in the error signal. In addition, it is known from the proof of Theorem~\ref{t1} that one necessary condition for achieving the lower bound in \eqref{bound9} is that $\mathbf{e}_k$ is with probability density function (corresponding to $p = \infty$)
\begin{flalign} &
f_{\mathbf{e}_{k}} \left( x \right)
= 
\left\{ \begin{array}{cc}
\frac{1}{2 \mu}, & \left| x \right| \leq \mu,\\
0, & \left| x \right| > \mu,
\end{array} \right.
& \end{flalign}
which represents a uniform distribution, indicating that the error should be steered to being with a uniform distribution.

\subsection{A Power-Spectral Characterization}

It follows directly from Theorem~\ref{t1} and \eqref{entropyrate} that if $\left\{ \mathbf{d}_k \right\}$ is asymptotically stationary, then
	\begin{flalign} \label{spectrum} & 
\limsup_{k \to \infty} \left[ \mathbb{E} \left( \left| \mathbf{e}_{k} \right|^{p} \right) \right]^{\frac{1}{p}} \nonumber \\
&~~~~ \geq \frac{1}{2 \Gamma \left( \frac{p+1}{p} \right) \left( p \mathrm{e} \right)^{\frac{1}{p}}} \left[\prod_{i=1}^{n} \max \left\{1, \left| \lambda_{i} \left( A \right) \right|  \right\} \right] 2^{ h_{\infty} \left( \mathbf{d} \right) }.
& \end{flalign}
In fact, for this particular case, a more specific formula could be derived in terms of power spectrum.

\begin{cor} \label{power1}
	Consider the control system given in Fig.~\ref{feedback}, where the plant $P$ is given by \eqref{plant} while the controller $K$ is given by \eqref{controller}. Let the disturbance $\left\{ \mathbf{d}_k \right\}$ be asymptotically stationary with asymptotic power spectrum $ S_{\mathbf{d}} \left( \omega \right)$. 
	If $K$ stabilizes $P$, then	
	\begin{flalign} \label{bound2}
	&\limsup_{k \to \infty} \left[ \mathbb{E} \left( \left| \mathbf{e}_{k} \right|^{p} \right) \right]^{\frac{1}{p}} \nonumber \\
	&~~~~ \geq \frac{\sqrt{ 2 \pi \mathrm{e}} }{2 \Gamma \left( \frac{p+1}{p} \right) \left( p \mathrm{e} \right)^{\frac{1}{p}}} \left[\prod_{i=1}^{n} \max \left\{1, \left| \lambda_{i} \left( A \right) \right|  \right\} \right] \nonumber \\ 
	&~~~~~~~~ \times \left[ 2^{- J_{\infty} \left( \mathbf{d} \right)} \right] 2^{\frac{1}{2 \mathrm{\pi}}\int_{-\mathrm{\pi}}^{\mathrm{\pi}}{\log \sqrt{S_{ \mathbf{d} } \left( \omega \right) } \mathrm{d}\omega }}.
	& \end{flalign}
	Herein, $S_{ \mathbf{d} } \left( \omega \right)$ denotes the asymptotic power spectrum of $\left\{ \mathbf{d}_k \right\}$, while $J_{\infty} \left( \mathbf{d} \right)$ denotes the negentropy rate \citep{fang2017towards} of $\left\{ \mathbf{d}_k \right\}$, whereas $J_{\infty} \left( \mathbf{d} \right) \geq 0$, and $J_{\infty} \left( \mathbf{d} \right) = 0$ if and only if $\left\{ \mathbf{d}_k \right\}$ is Gaussian.
\end{cor}

\begin{pf}
	It is known from \citet{fang2017towards} that for an asymptotically stationary stochastic process $\left\{ \mathbf{d}_k \right\}$ with asymptotic power spectrum $ S_{\mathbf{d}} \left( \omega \right)$, it holds that
	\begin{flalign} & 
	h_{\infty} \left( \mathbf{d} \right) 
	= \frac{1}{2 \mathrm{\pi}} \int_{-\mathrm{\pi}}^{\mathrm{\pi}} \log \sqrt{2 \pi \mathrm{e} S_{ \mathbf{d} } \left( \omega \right) } \mathrm{d}\omega - J_{\infty} \left( \mathbf{d} \right).  \nonumber
	& \end{flalign}
	Consequently,
	\begin{flalign}
	2^{h_{\infty} \left( \mathbf{d} \right)} 
	&= \left[ 2^{- J_{\infty} \left( \mathbf{d} \right)} \right] 2^{\frac{1}{2 \mathrm{\pi}}\int_{-\mathrm{\pi}}^{\mathrm{\pi}}{\log \sqrt{2 \pi \mathrm{e} S_{ \mathbf{d} } \left( \omega \right) } \mathrm{d}\omega }} \nonumber \\
	&= \sqrt{2 \pi \mathrm{e}} \left[ 2^{- J_{\infty} \left( \mathbf{d} \right)} \right] 2^{\frac{1}{2 \mathrm{\pi}}\int_{-\mathrm{\pi}}^{\mathrm{\pi}}{\log \sqrt{S_{ \mathbf{d} } \left( \omega \right) } \mathrm{d}\omega }}, \nonumber
	& \end{flalign}
	which, together with \eqref{spectrum}, completes the proof.
\qed \end{pf}

Herein, negentropy rate is a measure of non-Gaussianity for asymptotically stationary processes, which becomes smaller as the disturbance becomes more Gaussian; see, e.g., \citet{fang2017towards} for more details of its properties. Accordingly, the lower bound in \eqref{bound2} will increase as $\left\{ \mathbf{d}_k \right\}$ becomes more Gaussian, and vice versa. In the limit when  $\left\{ \mathbf{d}_k \right\}$ is Gaussian, \eqref{bound2} reduces to
\begin{flalign} \label{bound3}
&\limsup_{k \to \infty} \left[ \mathbb{E} \left( \left| \mathbf{e}_{k} \right|^{p} \right) \right]^{\frac{1}{p}} \nonumber \\
&~~~~ \geq \frac{\sqrt{ 2 \pi \mathrm{e}} }{2 \Gamma \left( \frac{p+1}{p} \right) \left( p \mathrm{e} \right)^{\frac{1}{p}}} \left[\prod_{i=1}^{n} \max \left\{1, \left| \lambda_{i} \left( A \right) \right|  \right\} \right] \nonumber \\
&~~~~~~~~ \times 2^{\frac{1}{2 \mathrm{\pi}}\int_{-\mathrm{\pi}}^{\mathrm{\pi}}{\log \sqrt{S_{ \mathbf{d} } \left( \omega \right) } \mathrm{d}\omega }}.
& \end{flalign}
In addition, if $A$ is further assumed to be stable, then \begin{flalign} &\prod_{i=1}^{n} \max \left\{1, \left| \lambda_{i} \left( A \right) \right|  \right\} = 1,
& \end{flalign} 
and \eqref{bound3} becomes
\begin{flalign} \label{bound4}
&\limsup_{k \to \infty} \left[ \mathbb{E} \left( \left| \mathbf{e}_{k} \right|^{p} \right) \right]^{\frac{1}{p}} \nonumber \\
&~~~~ \geq \frac{\sqrt{ 2 \pi \mathrm{e}} }{2 \Gamma \left( \frac{p+1}{p} \right) \left( p \mathrm{e} \right)^{\frac{1}{p}}} 2^{\frac{1}{2 \mathrm{\pi}}\int_{-\mathrm{\pi}}^{\mathrm{\pi}}{\log \sqrt{S_{ \mathbf{d} } \left( \omega \right) } \mathrm{d}\omega }},
& \end{flalign} 
which, in a broad sense, may be viewed as a generalization of the Kolmogorov--Szeg\"o formula (see, e.g., \citet{Pap:02, vaidyanathan2007theory, lindquist2015linear} and the references therein). More specifically, when $p = 2$, \eqref{bound4} reduces to
\begin{flalign} & \label{bound5}
\limsup_{k \to \infty} \left[ \mathbb{E} \left( \left| \mathbf{e}_{k} \right|^{2} \right) \right]^{\frac{1}{2}} 
\geq 2^{\frac{1}{2 \mathrm{\pi}}\int_{-\mathrm{\pi}}^{\mathrm{\pi}}{\log \sqrt{S_{ \mathbf{d} } \left( \omega \right) } \mathrm{d}\omega }},
& \end{flalign} 
which is equivalent to
\begin{flalign} &
\limsup_{k \to \infty}  \mathbb{E} \left(  \mathbf{e}_{k}^{2} \right)
\geq 2^{\frac{1}{2 \mathrm{\pi}}\int_{-\mathrm{\pi}}^{\mathrm{\pi}}{\log S_{ \mathbf{d} } \left( \omega \right) } \mathrm{d}\omega },
& \end{flalign} 
and coincides with the Kolmogorov--Szeg\"o formula; this indicates that the lower bound is tight in this particular case.

\subsection{From Plant Input to Plant Output}

In fact, the error signal considered in the previous subsection is essentially the plant input.
We now examine the $\mathcal{L}_p$ bounds on the plant output. Consider again the feedback control system depicted in Fig.~\ref{feedback}, where the plant is given by \eqref{plant} while the controller is given by \eqref{controller}. Let the integer $\nu > 0$ denote the relative degree of the plant's state-space model in \eqref{plant}; that is, $CA^{i}B = 0, \forall i = 0, \ldots, \nu -1$, whereas  $CA^{\nu}B \neq 0$. Denote $CA^{\nu}B = \rho$. Meanwhile, the finite zeros of the plant can be determined by the set
\begin{flalign} &
\mathcal{Z} = \left\{ x: \mathrm{rank} \begin{bmatrix}
A - x I & B \\
C & 0
\end{bmatrix}
< \max_{y \in \mathbb{C}} \mathrm{rank} \begin{bmatrix}
A - y I & B \\
C & 0
\end{bmatrix}
\right\}. \nonumber
& \end{flalign}

The following result characterizes the $\mathcal{L}_p$ norm of the plant output in terms of the nonminimum-phase zeros of the plant.

\begin{thm} \label{t2}
	Consider the control system depicted in Fig.~\ref{feedback}, where the plant $P$ is given by \eqref{plant} while the controller $K$ is given by \eqref{controller}. If $K$ stabilizes $P$, then the $\mathcal{L}_p$ norm of $\mathbf{y}_{k}$ is asymptotically lower bounded by 
	\begin{flalign} \label{boundy}
	&\limsup_{k \to \infty} 
	\left[ \mathbb{E} \left( \left| \mathbf{y}_{k} \right|^{p} \right) \right]^{\frac{1}{p}} \nonumber \\
	&~~~~ \geq \frac{\left| \rho \right|}{2 \Gamma \left( \frac{p+1}{p} \right) \left( p \mathrm{e} \right)^{\frac{1}{p}}} \left[\prod_{\varphi \in 
		\mathcal{Z}} \max \left\{1, \left| \varphi \right|  \right\} \right]
	\nonumber \\
	&~~~~~~~~ \times 2^{\limsup_{k \to \infty} h \left( \mathbf{d}_{k- \nu} |  \mathbf{d}_{0,\ldots,k- \nu-1} \right) },
	& \end{flalign}
	where $\prod_{\varphi \in 
		\mathcal{Z}} \max \left\{1, \left| \varphi \right|  \right\}$ essentially denotes the product of (the magnitudes of) all the nonminimum-phase zeros of the plant.
\end{thm}


\begin{pf}
	Note first that, according to Lemma~\ref{maximum}, we have
	\begin{flalign} &
	\left[ \mathbb{E} \left( \left| \mathbf{y}_{k} \right|^{p} \right) \right]^{\frac{1}{p}}
	\geq \frac{2^{h \left(  \mathbf{y}_{k} \right)}}{2 \Gamma \left( \frac{p+1}{p} \right) \left( p \mathrm{e} \right)^{\frac{1}{p}}}, \nonumber
	& \end{flalign}
	where equality holds
	if and only if $\mathbf{y}_{k}$ is with probability density function
	\begin{flalign} &
	f_{\mathbf{y}_{k}} \left( x \right)
	= \frac{ \mathrm{e}^{- \left| x \right|^{p} / \left( p \mu^{p} \right)} }{2 \Gamma \left( \frac{p+1}{p} \right) p^{\frac{1}{p}} \mu}, \nonumber
	& \end{flalign}
	whereas
	\begin{flalign} &
	\mu 
	= \frac{2^{h \left( \mathbf{y}_{k} \right)}}{2 \Gamma \left( \frac{p+1}{p} \right) \left( p \mathrm{e} \right)^{\frac{1}{p}}}. \nonumber
	& \end{flalign}
	Meanwhile, 
	\begin{flalign} 
	h \left( \mathbf{y}_{k} \right) 
	& = h \left( \mathbf{y}_{k } |  \mathbf{y}_{0,\ldots,k -1}, \mathbf{z}_{0}, \mathbf{x}_{0} \right) + I \left( \mathbf{y}_{k  }; \mathbf{y}_{0,\ldots,k-1}, \mathbf{z}_{0}, \mathbf{x}_{0} \right) \nonumber \\
	& = h \left( \mathbf{d}_{k - \nu} |  \mathbf{d}_{0,\ldots,k-\nu -1}, \mathbf{z}_{0}, \mathbf{x}_{0} \right) + \log \left| \rho \right| \nonumber \\
	&~~~~ + I \left( \mathbf{y}_{k  }; \mathbf{y}_{0,\ldots,k -1}, \mathbf{z}_{0}, \mathbf{x}_{0} \right), \nonumber
	& \end{flalign}
	since it is known from \citet{Oka:08} that
	\begin{flalign} 
	&h \left( \mathbf{y}_{k } |  \mathbf{y}_{0,\ldots,k -1}, \mathbf{z}_{0}, \mathbf{x}_{0} \right) \nonumber \\
	&~~~~ = h \left( \mathbf{d}_{k - \nu } |  \mathbf{d}_{0,\ldots,k-\nu -1}, \mathbf{z}_{0}, \mathbf{x}_{0} \right) + \log \left| \rho \right|. \nonumber
	& \end{flalign}
	On the other hand, it follows from the proof of Theorem~\ref{t1} that
	\begin{flalign} &  
	h \left( \mathbf{d}_{k - \nu} |  \mathbf{d}_{0,\ldots,k - \nu -1}, \mathbf{z}_{0}, \mathbf{x}_{0} \right) 
	= h \left( \mathbf{d}_{k - \nu} |  \mathbf{d}_{0,\ldots,k - \nu -1} \right). \nonumber
	& \end{flalign}
	As such,
	\begin{flalign} 
	h \left( \mathbf{y}_{k} \right)
	&= h \left( \mathbf{d}_{k - \nu} |  \mathbf{d}_{0,\ldots,k- \nu-1} \right) + \log \left| \rho \right| \nonumber \\
	&~~~~ + I \left( \mathbf{y}_{k}; \mathbf{y}_{0,\ldots,k-1}, \mathbf{z}_{0}, \mathbf{x}_{0} \right), \nonumber
	& \end{flalign}
	and thus
	\begin{flalign} 
	&\limsup_{k \to \infty} h \left( \mathbf{y}_{k} \right) \nonumber \\
	&= \limsup_{k \to \infty} \left[ h \left( \mathbf{d}_{k- \nu} |  \mathbf{d}_{0,\ldots,k- \nu-1} \right) +  I \left( \mathbf{y}_{k}; \mathbf{y}_{0,\ldots,k-1}, \mathbf{z}_{0}, \mathbf{x}_{0} \right) \right] 
	\nonumber \\
	&~~~~ + \log \left| \rho \right|
	\nonumber \\
	&\geq \limsup_{k \to \infty} h \left( \mathbf{d}_{k- \nu} |  \mathbf{d}_{0,\ldots,k- \nu-1} \right) \nonumber \\
	&~~~~ + \liminf_{k \to \infty}  I \left( \mathbf{y}_{k}; \mathbf{y}_{0,\ldots,k-1}, \mathbf{z}_{0}, \mathbf{x}_{0} \right) + \log \left| \rho \right|. \nonumber
	& \end{flalign}
	On the other hand,
	\begin{flalign} 
	&\liminf_{k \to \infty} I \left(\mathbf{y}_k; \mathbf{y}_{0,\ldots,k-1}, \mathbf{z}_0, \mathbf{x}_0 \right) \nonumber \\
	& = \liminf_{k \to \infty} \frac{I \left(\mathbf{y}_0; \mathbf{z}_0, \mathbf{x}_0)\right) + \cdots + I \left(\mathbf{y}_k; \mathbf{y}_{0,\ldots,k-1}, \mathbf{z}_0, \mathbf{x}_0\right)}{k+1}. \nonumber
	& \end{flalign}
	Meanwhile,
	\begin{flalign}
	& I \left( \mathbf{y}_k; \mathbf{y}_{0,\ldots,k-1}, \mathbf{z}_0, \mathbf{x}_0 \right) \nonumber \\
	& = I \left( \mathbf{y}_k; \mathbf{y}_{0,\ldots,k-1}, \mathbf{x}_0 \right) +  I \left( \mathbf{y}_k; \mathbf{z}_0 | \mathbf{y}_{0,\ldots,k-1}, \mathbf{x}_0 \right) \nonumber \\
	& \geq  I \left( \mathbf{y}_k; \mathbf{y}_{0,\ldots,k-1}, \mathbf{x}_0 \right), \nonumber
	& \end{flalign}
	while
	\begin{flalign}
	&I \left(\mathbf{y}_k; \mathbf{y}_{0,\ldots,k-1}, \mathbf{x}_0\right) \nonumber \\ 
	&~~~~ = I \left(\mathbf{y}_k; \mathbf{x}_0|\mathbf{y}_{0,\ldots,k-1}\right) + I \left(\mathbf{y}_k; \mathbf{y}_{0,\ldots,k-1}\right) \nonumber \\ 
	&~~~~ \geq I \left(\mathbf{y}_k; \mathbf{x}_0|\mathbf{y}_{0,\ldots,k-1}\right).\nonumber
	& \end{flalign}
	As a result,
	\begin{flalign}
	&I \left(\mathbf{y}_k; \mathbf{y}_{0,\ldots,k-1}, \mathbf{z}_0, \mathbf{x}_0\right) \geq I \left(\mathbf{y}_k; \mathbf{x}_0|\mathbf{y}_{0,\ldots,k-1}\right),\nonumber
	& \end{flalign}
	or more specifically,
	\begin{flalign} &
	\left\{ \begin{array}{rcl}
	I \left(\mathbf{y}_0; \mathbf{z}_0, \mathbf{x}_0\right) & \geq & I \left( \mathbf{y}_{0}; \mathbf{x}_{0} \right),\\
	I \left(\mathbf{y}_1; \mathbf{y}_{0}, \mathbf{z}_0, \mathbf{x}_0\right)& \geq & I \left( \mathbf{y}_{1}; \mathbf{x}_{0} | \mathbf{y}_{0} \right),
	\\
	& \vdots & 
	\\
	I \left(\mathbf{y}_k; \mathbf{y}_{0,\ldots,k-1}, \mathbf{z}_0, \mathbf{x}_0\right) & \geq & I \left( \mathbf{y}_{k}; \mathbf{x}_{0} | \mathbf{y}_{0,\ldots,k-1} \right).
	\end{array} \right. \nonumber
	& \end{flalign}
	Accordingly,
	\begin{flalign} 
	&I \left(\mathbf{y}_0; \mathbf{z}_0, \mathbf{x}_0\right) + I \left(\mathbf{y}_1; \mathbf{y}_0, \mathbf{z}_0, \mathbf{x}_0\right) + \cdots \nonumber \\
	&~~~~+ I \left(\mathbf{y}_k; \mathbf{y}_{0,\ldots,k-1}, \mathbf{z}_0, \mathbf{x}_0\right) \nonumber \\
	& \geq I \left( \mathbf{y}_{0}; \mathbf{x}_{0} \right) + I \left( \mathbf{y}_{1}; \mathbf{x}_{0} | \mathbf{y}_{0} \right) + \cdots + I \left( \mathbf{y}_{k}; \mathbf{x}_{0} | \mathbf{y}_{0,\ldots,k-1} \right) \nonumber \\
	& = I \left( \mathbf{y}_{0, \ldots, k}; \mathbf{x}_{0} \right). \nonumber
	& \end{flalign}
	Furthermore, it is known from \citet{Oka:08} that
	\begin{flalign} & 
	\liminf_{k \to \infty} \frac{I \left( \mathbf{y}_{0, \ldots, k}; \mathbf{x}_{0} \right)}{k+1} \geq \sum_{\varphi \in 
		\mathcal{Z}} \max \left\{0, \log \left| \varphi \right|  \right\}. \nonumber
	& \end{flalign}
	Hence,
	\begin{flalign} 
	\limsup_{k \to \infty} h \left( \mathbf{y}_{k} \right)
	&\geq \limsup_{k \to \infty} h \left( \mathbf{d}_{k- \nu} |  \mathbf{d}_{0,\ldots,k- \nu-1} \right) + \log \left| \rho \right| \nonumber \\
	&~~~~ + \sum_{\varphi \in 
		\mathcal{Z}} \max \left\{0, \log \left| \varphi \right|  \right\}. \nonumber
	& \end{flalign}
	Accordingly,
	\begin{flalign} 
	&\limsup_{k \to \infty} 
	\left[ \mathbb{E} \left( \left| \mathbf{y}_{k} \right|^{p} \right) \right]^{\frac{1}{p}}
	\geq \limsup_{k \to \infty} \frac{2^{h \left(  \mathbf{y}_{k} \right)}}{2 \Gamma \left( \frac{p+1}{p} \right) \left( p \mathrm{e} \right)^{\frac{1}{p}}} \nonumber \\
	& = \frac{1}{2 \Gamma \left( \frac{p+1}{p} \right) \left( p \mathrm{e} \right)^{\frac{1}{p}}} 2^{\limsup_{k \to \infty} h \left(  \mathbf{y}_{k} \right) } \nonumber \\
	& \geq \frac{1}{2 \Gamma \left( \frac{p+1}{p} \right) \left( p \mathrm{e} \right)^{\frac{1}{p}}} \nonumber \\
	&\times 2^{\limsup_{k \to \infty} h \left( \mathbf{d}_{k- \nu} |  \mathbf{d}_{0,\ldots,k- \nu-1} \right) + \log \left| \rho \right| + \sum_{\varphi \in 
			\mathcal{Z}} \max \left\{0, \log \left| \varphi \right|  \right\}}
	\nonumber \\
	& = \frac{\left| \rho \right|}{2 \Gamma \left( \frac{p+1}{p} \right) \left( p \mathrm{e} \right)^{\frac{1}{p}}} \left[\prod_{\varphi \in 
		\mathcal{Z}} \max \left\{1, \left| \varphi \right|  \right\} \right]
	\nonumber \\
	&~~~~ \times 2^{\limsup_{k \to \infty} h \left( \mathbf{d}_{k- \nu} |  \mathbf{d}_{0,\ldots,k- \nu-1} \right) }. \nonumber
	& \end{flalign}
	This completes the proof.
\qed \end{pf}

Note that further interpretations and implications of Theorem~\ref{t2} can be discussed in a similar manner to those for Theorem~\ref{t1} as well. For instance, if $\left\{ \mathbf{d}_{k} \right\}$ is assumed to be asymptotically stationary, then it holds that \citep{Cov:06}
\begin{flalign} 
\limsup_{k \to \infty} h \left( \mathbf{d}_{k- \nu} |  \mathbf{d}_{0,\ldots,k- \nu-1} \right)
&= \lim_{k \to \infty} h \left( \mathbf{d}_{k- \nu} |  \mathbf{d}_{0,\ldots,k- \nu-1} \right) \nonumber \\
&= h_{\infty} \left( \mathbf{d} \right).
& \end{flalign}
and \eqref{boundy} becomes
\begin{flalign} 
&\limsup_{k \to \infty} 
\left[ \mathbb{E} \left( \left| \mathbf{y}_{k} \right|^{p} \right) \right]^{\frac{1}{p}} \nonumber \\
&~~~~ \geq \frac{\left| \rho \right|}{2 \Gamma \left( \frac{p+1}{p} \right) \left( p \mathrm{e} \right)^{\frac{1}{p}}} \left[\prod_{\varphi \in 
	\mathcal{Z}} \max \left\{1, \left| \varphi \right|  \right\} \right] 2^{ h_{\infty} \left( \mathbf{d} \right) },
& \end{flalign}
which may in turn be analyzed using a power-spectral characterization as well.

\subsection{From LTI Plants to Generic Plants}

Finally, we investigate the case when the plant is also generically assumed to be (strictly) causal. More specifically, consider the feedback system in Fig.~\ref{feedback}, where $\mathbf{d}_k, \mathbf{e}_k, \mathbf{y}_k, \mathbf{z}_k \in \mathbb{R}$. In this generic setting, for the plant we assume that for any time instant $k \geq 0$, 
\begin{flalign} & \label{generic1}
\mathbf{y}_k = P_{k} \left( \mathbf{e}_{0,\ldots,k-1} \right), 
& \end{flalign}
where $P_{k} \left( \cdot \right)$ may represent any deterministic or randomized functions/mappings. Meanwhile, the controller is still generically assumed to be causal as
\begin{flalign} & \label{generic2}
\mathbf{z}_k = K_{k} \left( \mathbf{y}_{0,\ldots,k} \right),
& \end{flalign}
where $K_{k} \left( \cdot \right)$ may represent any deterministic or randomized functions/mappings.
In addition, $\left\{ \mathbf{d}_k \right\}$, $\mathbf{y}_0$, and $\mathbf{z}_0$ are assumed to be mutually independent.

The following result provides a lower bound on the $\mathcal{L}_p$ norm of the error signal for this generic setting.

\begin{thm} \label{t3}
	Consider the control system depicted in Fig.~\ref{feedback}, where the plant $P$  is given by \eqref{generic1} while the controller $K$ is given by \eqref{generic2}. Then, the $\mathcal{L}_p$ norm of $ \mathbf{e}_{k} $ is lower bounded by 
	\begin{flalign} &
	\left[ \mathbb{E} \left( \left| \mathbf{e}_{k} \right|^{p} \right) \right]^{\frac{1}{p}}
	\geq \frac{2^{h \left( \mathbf{d}_k | \mathbf{d}_{0,\ldots,k-1} \right)}}{2 \Gamma \left( \frac{p+1}{p} \right) \left( p \mathrm{e} \right)^{\frac{1}{p}}}. 
	& \end{flalign}
	In the asymptotic case, it holds that
	\begin{flalign} \label{bound7}
	&\limsup_{k \to \infty} 
	\left[ \mathbb{E} \left( \left| \mathbf{e}_{k} \right|^{p} \right) \right]^{\frac{1}{p}} \nonumber \\
	&~~~~ \geq \frac{1}{2 \Gamma \left( \frac{p+1}{p} \right) \left( p \mathrm{e} \right)^{\frac{1}{p}}} 2^{\limsup_{k \to \infty} h \left( \mathbf{d}_{k} |  \mathbf{d}_{0,\ldots,k-1} \right) }.
	& \end{flalign}
\end{thm}


\begin{pf}
	To begin with, it follows from Lemma~\ref{maximum} that
	\begin{flalign} & 
	\left[ \mathbb{E} \left( \left| \mathbf{e}_{k} \right|^{p} \right) \right]^{\frac{1}{p}}
	\geq \frac{2^{h \left(  \mathbf{e}_{k} \right)}}{2 \Gamma \left( \frac{p+1}{p} \right) \left( p \mathrm{e} \right)^{\frac{1}{p}}}, \nonumber
	& \end{flalign}
	where equality holds if and only if $\mathbf{e}_{k}$ is with probability density function
	\begin{flalign} &
	f_{\mathbf{e}_{k}} \left( x \right)
	= \frac{ \mathrm{e}^{- \left| x \right|^{p} / \left( p \mu^{p} \right)} }{2 \Gamma \left( \frac{p+1}{p} \right) p^{\frac{1}{p}} \mu}, \nonumber
	& \end{flalign}	
	whereas
	\begin{flalign} &
	\mu 
	= \frac{2^{h \left( \mathbf{e}_{k} \right)}}{2 \Gamma \left( \frac{p+1}{p} \right) \left( p \mathrm{e} \right)^{\frac{1}{p}}}. \nonumber
	& \end{flalign}

	On the other hand, we will prove the fact that $\mathbf{z}_{k}$ is eventually a function of $\mathbf{d}_{0,\ldots,k-1}$, $\mathbf{y}_{0}$,  and $\mathbf{z}_{0}$.
	More specifically, when $k=1$, it follows from \eqref{generic1} and \eqref{generic2} that
	\begin{flalign} 
	\mathbf{z}_{1} &= K_{1} \left( \mathbf{y}_{0}, \mathbf{y}_{1} \right) = K_{1} \left( \mathbf{y}_{0}, P_{1} \left( \mathbf{e}_{0} \right) \right) \nonumber \\
	&= K_{1} \left( \mathbf{y}_{0}, P_{1} \left( \mathbf{d}_{0} + \mathbf{z}_{0} \right) \right), \nonumber
	& \end{flalign}
	that is, $\mathbf{z}_{1}$ is a function of $\mathbf{d}_{0}$, $\mathbf{y}_{0}$, and $\mathbf{z}_{0}$.
	Next, when $k=2$, it follows from
	\eqref{generic1} and \eqref{generic2} that
	\begin{flalign} 
	\mathbf{z}_{2} & = K_{2} \left( \mathbf{y}_{0}, \mathbf{y}_{1} , \mathbf{y}_{2} \right) = K_{2} \left( \mathbf{y}_{0}, P_{1} \left( \mathbf{e}_{0} \right), P_{2} \left( \mathbf{e}_{0}, \mathbf{e}_{1} \right) \right) 
	\nonumber \\ 
	&= K_{2} \left( \mathbf{y}_{0}, P_{1} \left( \mathbf{d}_{0} + \mathbf{z}_{0} \right), P_{2} \left( \mathbf{d}_{0} + \mathbf{z}_{0}, \mathbf{d}_{1} + \mathbf{z}_{1} \right) \right). 
	\nonumber
	& \end{flalign}
	As such, noting also that
	\begin{flalign}
	& \mathbf{z}_{1} = K_{1} \left( \mathbf{y}_{0}, P_{1} \left( \mathbf{d}_{0} + \mathbf{z}_{0} \right) \right), \nonumber &
	\end{flalign}
	it is clear that
	$\mathbf{z}_{2}$ is a function of $\mathbf{d}_{0,1}$, $\mathbf{y}_{0}$, and $\mathbf{z}_{0}$. We may then repeat this process and show that for any $k \geq 0$, $\mathbf{z}_{k}$ is eventually a function of $\mathbf{d}_{0,\ldots,k-1}$, $\mathbf{y}_{0}$,  and $\mathbf{z}_{0}$.
	
	We will then proceed to prove the main result of this theorem. Note first that
	\begin{flalign} 
	&h \left( \mathbf{e}_{k} \right)
	= h \left( \mathbf{e}_{k} |  \mathbf{d}_{0,\ldots,k-1}, \mathbf{y}_{0}, \mathbf{z}_{0} \right) + I \left( \mathbf{e}_{k}; \mathbf{d}_{0,\ldots,k-1}, \mathbf{y}_{0}, \mathbf{z}_{0} \right) \nonumber \\
	& = h \left( \mathbf{z}_{k} + \mathbf{d}_{k} |  \mathbf{d}_{0,\ldots,k-1}, \mathbf{y}_{0}, \mathbf{z}_{0} \right) + I \left( \mathbf{e}_{k}; \mathbf{d}_{0,\ldots,k-1}, \mathbf{y}_{0}, \mathbf{z}_{0} \right). \nonumber
	& \end{flalign}
	Then, according to the fact that $\mathbf{z}_{k}$ is a function of $\mathbf{d}_{0,\ldots,k-1}$, $\mathbf{y}_{0}$, and $\mathbf{z}_{0}$, we have
	\begin{flalign} & 
	h \left( \mathbf{z}_{k} + \mathbf{d}_{k} |  \mathbf{d}_{0,\ldots,k-1}, \mathbf{y}_{0}, \mathbf{z}_{0} \right) = h \left( \mathbf{d}_{k} |  \mathbf{d}_{0,\ldots,k-1}, \mathbf{y}_{0}, \mathbf{z}_{0} \right). \nonumber
	& \end{flalign}
	On the other hand, since $ \left\{ \mathbf{d}_{k} \right\}$, $\mathbf{y}_{0}$, and $\mathbf{z}_{0}$ are mutually independent (and thus $\mathbf{d}_{k}$ is independent $\mathbf{y}_{0}$ and $\mathbf{z}_{0}$ given $\mathbf{d}_{0,\ldots,k-1}$), we have
	\begin{flalign} 
	& h \left( \mathbf{d}_{k} |  \mathbf{d}_{0,\ldots,k-1}, \mathbf{y}_{0}, \mathbf{z}_{0} \right) \nonumber \\
	&~~~~ = h \left( \mathbf{d}_{k} |  \mathbf{d}_{0,\ldots,k-1} \right)  - I \left( \mathbf{d}_{k}; \mathbf{y}_{0}, \mathbf{z}_{0} |  \mathbf{d}_{0,\ldots,k-1} \right) \nonumber \\
	&~~~~ = h \left( \mathbf{d}_{k} |  \mathbf{d}_{0,\ldots,k-1} \right). \nonumber
	& \end{flalign}
	As a result,
	\begin{flalign} & 
	h \left( \mathbf{e}_{k} \right)
	= h \left( \mathbf{d}_{k} |  \mathbf{d}_{0,\ldots,k-1} \right) + I \left( \mathbf{e}_{k}; \mathbf{d}_{0,\ldots,k-1}, \mathbf{y}_{0}, \mathbf{z}_{0} \right). \nonumber
	& \end{flalign}
	Hence,
	\begin{flalign} & 
	2^{ h \left( \mathbf{e}_{k} \right)} 
	\geq 2^{h \left( \mathbf{d}_k | \mathbf{d}_{0,\ldots,k-1} \right)}, \nonumber
	& \end{flalign}
	and
	\begin{flalign} &
	\left[ \mathbb{E} \left( \left| \mathbf{e}_{k} \right|^{p} \right) \right]^{\frac{1}{p}}
	\geq \frac{2^{h \left( \mathbf{d}_k | \mathbf{d}_{0,\ldots,k-1} \right)}}{2 \Gamma \left( \frac{p+1}{p} \right) \left( p \mathrm{e} \right)^{\frac{1}{p}}}. \nonumber
	& \end{flalign}
	In addition,
	\begin{flalign} &
	\limsup_{k \to \infty} \left[ \mathbb{E} \left( \left| \mathbf{e}_{k} \right|^{p} \right) \right]^{\frac{1}{p}} \nonumber \\
	&~~~~ \geq \frac{1}{2 \Gamma \left( \frac{p+1}{p} \right) \left( p \mathrm{e} \right)^{\frac{1}{p}}} 2^{ \limsup_{k \to \infty} h \left( \mathbf{d}_k | \mathbf{d}_{0,\ldots,k-1} \right)}. \nonumber
	& \end{flalign}
    This completes the proof.
\qed \end{pf}

Note that Theorem~\ref{t3} provides lower bounds for both the non-asymptotic case and asymptotic case. 
Meanwhile, such bounds hold as long as the plant is (strictly) causal, whether it be linear or nonlinear, time-invariant or time-varying, and so on. On the other hand, when the plant is LTI, the lower bound obtained in Theorem~\ref{t1} will then be tighter than that of \eqref{bound7} in general since
\begin{flalign} &
\prod_{i=1}^{n} \max \left\{1, \left| \lambda_{i} \left( A \right) \right|  \right\} \geq 1.
& \end{flalign}

\subsection{Generality of the Performance Bounds} \label{loop}


Note that for the fundamental $\mathcal{L}_p$ bounds derived in this paper, the classes of control algorithms that can be applied are not restricted in general, as long as they are causal and stabilizing. This means that the performance bounds are valid for all possible control design methods in practical use, including conventional methods as well as machine learning approaches such as reinforcement learning and deep learning (see, e.g., \citet{lewis2012reinforcement, mnih2015human, duan2016benchmarking, kocijan2016modelling, duriez2017machine, recht2019tour, bertsekas2019reinforcement, tiumentsev2019neural, zoppoli2020neural, hardt2021patterns} and the references therein). In particular, note that any machine learning algorithms in the position of the controller  can be viewed as causal (deterministic or randomized) functions/mappings from the controller input to the controller output, no matter what the specific algorithms are or how the parameters are to be tuned. As such, the aforementioned fundamental limitations are still valid with any learning elements in the feedback loop, that is to say, fundamental limits in general exist to what learning algorithms can achieve in the position of the controller, featuring fundamental limits of learning-based control; cf. also discussions on ``The limits of learning in feedback loops" in Chapter~12 of \citet{hardt2021patterns}. Meanwhile, note that, for instance, it is true that multilayer feedforward neural networks are universal approximators \citep{cybenko1989approximation, hornik1989multilayer, goodfellow2016deep}, but it is also true that the performance bounds hold for any functions the neural networks might approximate.


\section{Conclusion}

In this paper, we have presented the fundamental $\mathcal{L}_p$ bounds when controlling stochastic dynamical systems, which hold for any causal (stabilizing) controllers and any stochastic disturbances. We have considered both the case of LTI plants and the case of (strictly) causal plants. We have also provided discussions on the implications and the generality of the lower bounds.

Potential future research directions include investigating further the tightness of the derived bounds as well as how to achieve/approach them. It might also be interesting to examine the implications of the generic bounds in state estimation systems.

\balance

\bibliographystyle{apalike}
\bibliography{references}

\end{document}